# On the Performance of Joint Fingerprint Embedding and Decryption Scheme

Shiguo Lian, Zhongxuan Liu, Zhen Ren, Haila Wang

France Telecom R&D Beijing, Beijing, 100080, P.R China
E-mail: sg_lian@163.com

**Abstract.** Till now, few work has been done to analyze the performances of joint fingerprint embedding and decryption schemes. In this paper, the security of the joint fingerprint embedding and decryption scheme proposed by Kundur et al. is analyzed and improved. The analyses include the security against unauthorized customer, the security against authorized customer, the relationship between security and robustness, the relationship between security and imperceptibility and the perceptual security. Based these analyses, some means are proposed to strengthen the system, such as multi-key encryption and DC coefficient encryption. The method can be used to analyze other JFD schemes. It is expected to provide valuable information to design JFD schemes.

## 1 Introduction

Secure multimedia distribution becomes more and more important with the wide application of multimedia data and the fast development of networks. Generally, encryption method [1,2] is used to protect media data's confidentiality. Thus, only the authorized receiver can decrypt the data stream correctly. However, encryption method cannot solve the Super Distribution problem [3], that is, the decrypted copies can be distributed freely. To solve this problem, the fingerprint-based scheme [4,5,6,7] has been proposed. That is, for each customer, a unique watermark (fingerprint) is embedded into the multimedia program. Thus, each customer receives a different copy, and the fingerprint identifies the customer. This scheme can trace the illegal redistributors who send their copies to unauthorized customers.

One of the difficulties in fingerprint-based scheme is how to distribute the fingerprinted copies efficiently. Generally, there are three methods: embed fingerprint at the server end, in the router or at the receiver end. Straightforwardly, the server could embed a fingerprint in the media data and then send it to the according customer. However, considering that many customers may ask for the service at the same time, it is not practical for the server to send different copies to different customers efficiently. The second scheme, named Water-Casting, embeds watermark by the sub-servers [6], which distributes the server's loading to the sub-servers. However, the transmission protocol should be modified, which is not compliant with network protocol. Another scheme [7] embeds a fingerprint at the receiver end. Thus, the server sends only one media stream to all the customers. The potential difficulty is to keep the security of the operation at the customer end.

Embedding the fingerprint at the customer end may be practical if the security can be confirmed. Generally, media data are encrypted during transmission. Thus, at the receiver end, media data should be decrypted before displayed. If the fingerprint is embedded after the media data are decrypted, the decrypted media data may be leaked out before fingerprint embedding process. As an alternative, the joint fingerprint embedding and decryption (JFD) scheme [8,9] decrypts and embeds the media data at the same time, which avoids the leakage of the decrypted media data.

For JFD schemes, the security and robustness are two important performances. Till now, some methods have been proposed, e.g., Chamleon scheme [8], Kundur et al's scheme [9], Lian et al's scheme [10] and Lemma et al's scheme [11]. Chamleon scheme based on a stream cipher encrypts the media data with a stream cipher, decrypts and fingerprints the media data by modifying the LSB bits. This scheme is secure in cryptographic aspects [8], but is not robust to signal processing, such as recompression, adding noise, etc. The scheme proposed by Lian et al [10] encrypts media data at the server side by encrypting the variable-length code's index, and decrypts media data at the customer side by recovering code's index with both decryption and fingerprinting. This scheme is security against cryptographic attacks [8], while the robustness against general operations can not be confirmed. The scheme proposed by Lemma et al [11] encrypts media data by a key stream, and decrypts media data with a new key stream. It uses two different key streams for encryption and decryption respectively, which is similar with Chamleon method [8]. The scheme costs much time and space to transmit the decryption key stream. Kundur et al's scheme [9] is based on partial encryption, which confuses the sign bits of the DCT coefficients in encryption and decrypts only part of the sign bits in decryption. The position of the left sign bits determines the fingerprint. This scheme is often robust to some signal processing operations. Compared with previous schemes, Kundur et al's scheme is more suitable for compressed data. However, considering that most of the quantized coefficients are zeros, the method's security should be further discussed. Additionally, the number of the coefficients to be encrypted is in close relation with the robustness and imperceptibility.

Till now, few work has been done to discuss the performances of the JFD schemes. In this paper, taking Kundur et al's scheme for example, the performances of the JFD scheme are investigated and analyzed. We try to give a detailed discussion on its security, point out the relationship between its security and its robustness, imperceptibility, and propose some means to improve its performances. It is expected to provide valuable advices to design a secure JFD scheme. The rest of the paper is arranged as follows. In Section 2, Kundur et al's scheme is briefly introduced. The cryptographic security, relation between security and robustness, relation between security and imperceptibility and perceptual security are analyzed in Section 3, 4, 5 and 6, respectively. In Section 7, some means are proposed to improve the scheme's performances. Finally, conclusions are drawn, and future work is given in Section 8.

## 2 The JFD Scheme Proposed by Kundur et al

### 2.1 Introduction to Kundur et al's Scheme

Kundur et al proposed the joint fingerprint embedding and decryption scheme based on partial encryption [9]. In encryption, the DCT coefficients' signs are encrypted, while the amplitudes are left unchanged. In decryption, some of the signs are decrypted, while the others are left unchanged. Different key produces the copy with different signs left unchanged. The position of the left signs determines the uniqueness of the decrypted media data. The process is shown in Fig. 1 where the DCT coefficients' signs are confused by sign encryption under the control of $K_E$, and are decrypted into different copy under the control of $K_{D,j}$ (j=0,1,…,M-1) (M is the number of customers).

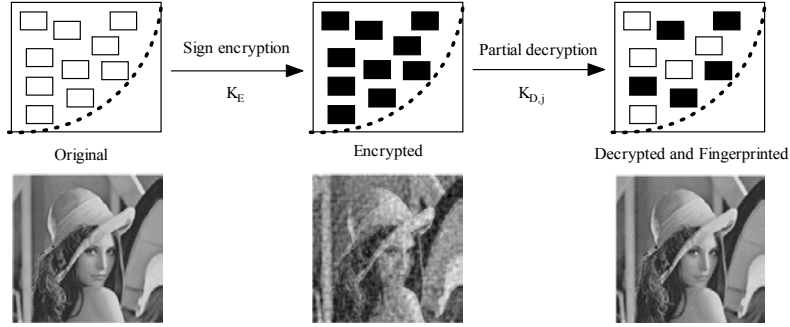

**Fig. 1.** The Architecture of Kundur et al's Scheme

### 2.2 Alternative Implementation of the Proposed JFD Scheme

Considering that the coefficients in the low frequency of DCT blocks are sensitive to images' intelligibility, they should be decrypted at the receiver end, which has not been considered in [9]. In the following content, we tend to leave only some coefficients in the middle or high frequency undecrypted. For convenience, some parameters are set: $N$, $N_T$ and $N_{nonzero}$. Among them, $N$ denotes the total number of the DCT coefficient, $N_{nonzero}$ ($0<N_{nonzero}\leq N$) denotes the number of nonzero coefficients that are actually encrypted at the server end while $N_T$ ($0<N_T\leq N_{nonzero}$) denotes the number of coefficients that are decrypted at the receiver end. Thus, the N coefficients are operated according to the following steps (as shown in Fig. 2):

  i) Zigzag scan: N coefficients are first ordered in zigzag mode $x_0 x_1 \ldots x_{N-1}$;

  ii) Sign encryption: The first Nnonzero coefficients $x_0 x_1 \ldots x_{Nnonzero-1}$ are encrypted;

  iii) Partial decryption: The first $N_T$ coefficients $x_0 x_1 \ldots x_{Nt-1}$ are decrypted, while the left $N_{nonzero}-N_T$ coefficients are partially decrypted;

  iv) Inv-zigzag scan: The processed coefficients are set back to the DCT block in inv-zigzag order.

## 3 The Cryptographic Security

This JFD scheme is first an encryption scheme, and also a watermarking or fingerprinting scheme. As an encryption scheme, the cryptographic security should be analyzed. Different from traditional encryption schemes, two attack cases arise in Kundur's scheme. One case is that the unauthorized customer tries to obtain the plain-media, which is named unauthorized attack by us. The other case is that the authorized customer tries to obtain another plain-media, which is named authorized attack by us. The security against both the attacks is presented as follows, respectively.

  *Security against Unauthorized Attack* In unauthorized attacks, the authorized customers have no keys to decrypt the media data. To get the plain-media is attractive to them. The direct method is the brute-force method. Generally, the brute-force space of a cryptosystem is the key space. However, in Kundur's scheme, the case is different. That is because the one encryption key corresponds to several decryption keys. Set the encryption space of $K_E$ be S (S is the key's space, for example $S=2^{64}$ if the key is of 64-bit length). The relation between the encryption key and the decryption key is shown in Fig. 3(a). Here, for each encryption key $K_E$, M decryption keys $K_{D,0}$, $K_{D,1}$, …, $K_{D,M-1}$ can be used to decrypt the media data into an intelligible copy. Thus, the brute-force space is reduced to

$$f_{un}(M) = S - M + 1.$$

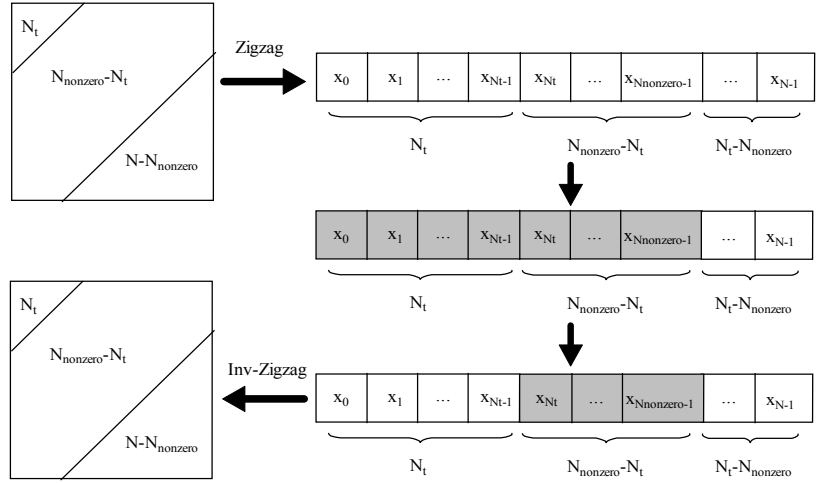

**Fig. 2.** The Encryption/Decryption Process in Detail

That is, an unauthorized customer needs only to try S-M+1 times but not S times before he can obtain the plain-media. Here, the space is a function of the customer number M under certain S. The relation is shown in Fig. 3(b) where $S=2^{64}$. As can be seen, under certain key space S, the brute-force space decreases with the rise of the customer number M. Thus, in order to keep secure, the key space S should be big enough, otherwise, the customer number M should be keep small enough.

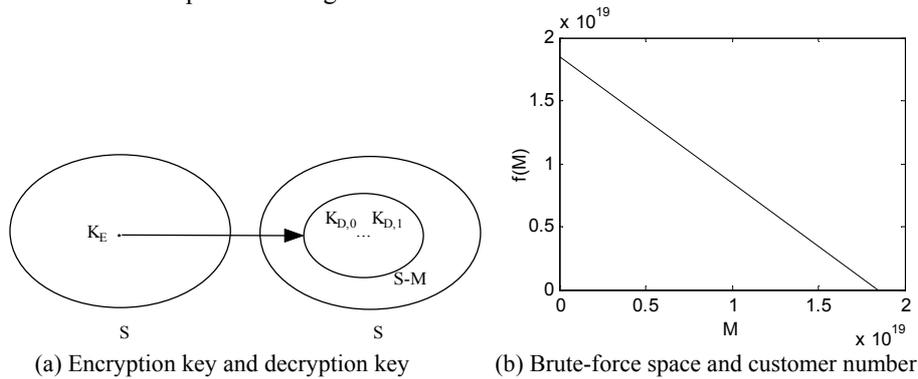

(a) Encryption key and decryption key  (b) Brute-force space and customer number

**Fig. 3.** Brute-force Space under Unauthorized Attacks

*Security against Authorized Attack* In authorized attacks, the authorized customer can decrypt the media data into a plain copy, but he doesn't want to distribute his copy to others. He may attack the scheme by brute-force, and obtain a plain-copy different from his own. In this case, the brute-force space is

$$f_{au}(M) = S - M .$$

For several authorized customers, they may produce a new copy through collusion operations, such as averaging or linear addition [12,13]. This case belongs to the domain of fingerprint encoding [14,15] that will not be discussed here.

## 4 The Relationship between Security and Robustness

In Kundur's scheme, the signs of DCT coefficients are encrypted, and the signs' position determines the uniqueness. In the aspect of security, the more the number of the encrypted signs, the more secure the scheme is. Differently, in the aspect of robustness, the number of the encrypted signs should be carefully decided. Generally, after quantization or re-quantization, more coefficients become zeros, thus the signs are changed, which affect the fingerprint detection. Thus, there are relation between security and robustness. Taking quantization for example, we discuss the relation in the following content.

*Security and Nonzero Coefficient* In this scheme, only the signs of the coefficients are encrypted. Thus, the number of the nonzero coefficients determines the encryption space. Set the number of the nonzero coefficients be $N_{nonzero}$. The encryption space of sign encryption is

$$S_{sign} = 2^{N_{nonzero}} .$$

*Quantization and $N_{nonzero}$* In quantization, the quantization factor q determines the number of nonzero coefficients. Generally, the bigger the quantization factor q is, the smaller $N_{nonzero}$ is. That is, $N_{nonzero}$ and q satisfy $N_{nonzero} = o(1/q)$. Fig. 4(a) shows the statistical results. Here, the quantization table in JPEG [16] is adopted, and q ranges from 0.1 to 3. As can be seen, $N_{nonzero}$ decreases with the rise of q.

*Security and Quantization* The relation between security and quantization satisfies

$$S_{sign} = 2^{o\left(\frac{1}{q}\right)} .$$

Thus, the encryption space decreases with the rise of the quantization factor. The relation is shown in Fig. 4(b). That is, the encryption system obtains higher security when q is small. Otherwise, if the media data are greatly quantized, few nonzero coefficients are left, and the scheme is of lower security.

## 5 The Relationship between Security and Imperceptibility

In Kundur et al's scheme, not all the signs of the coefficients can be left unencrypted. In order to keep the fingerprinted media data imperceptible, only the coefficients that are not sensitive to perceptual quality can be left. Thus, except zero coefficients and the insensitive ones, the remaining coefficients can be used to fingerprint the media data. The number of the remaining coefficients determines the number of the customers the scheme can support. Thus, there is relation between security and imperceptibility, which is presented as follows.

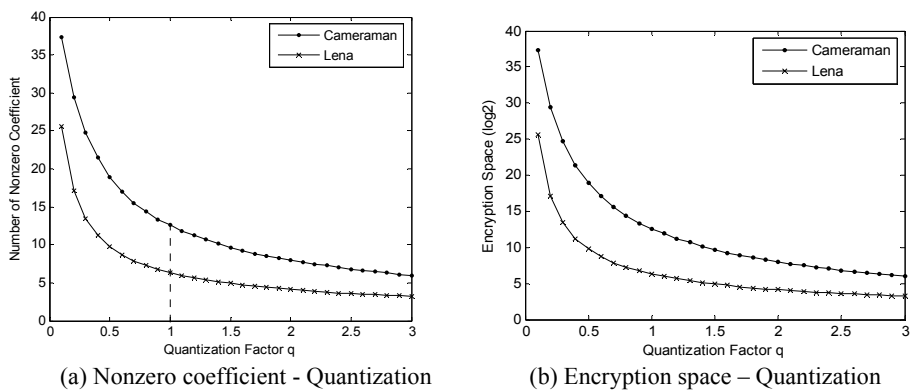

(a) Nonzero coefficient - Quantization    (b) Encryption space – Quantization

**Fig. 4.** Relation between robustness and security

*Imperceptibility and Threshold Coefficient* According to the distribution property of DCT coefficients, the sensitivity decreases with the rise of the coefficient's frequency, as shown in Fig. 5. Thus, the threshold coefficient $N_T$ is defined, which denotes the threshold position, from which to higher frequency, the coefficients can be fingerprinted. Generally, the imperceptibility I and threshold coefficient $N_T$ satisfy $N_T = o(I)$. That is, the higher the imperceptibility is required, the bigger the threshold coefficient $N_T$ should be. Fig. 7 shows the sensitivity of different coefficient, the relation between the threshold coefficient $N_T$ and the number of nonzero coefficients $N_{nonzero}$. As can be seen, the coefficient's sensitivity decreases with the rise of coefficients. In order to keep the PSNR no lower than 55, the threshold $N_T$ should be no smaller than 7. Additionally, $N_T < N_{nonzero}$ should be satisfied.

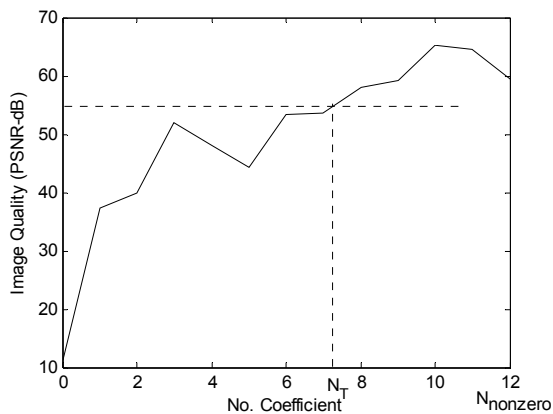

**Fig. 5.** Coefficient Sensitivity in DCT Transformation

*Security and Threshold Coefficient* In Kundur's scheme, the first $N_T$ coefficients should be decrypted completely, while the remained $N_{nonzero}$-$N_T$ ones are fingerprinted. Thus, the number of customers is determined by

$$M = 2^{N_{nonzero}-N_T}.$$

The bigger $N_T$ is, the fewer the customers can be supported. Thus, according to (1), the brute-force space is

$$f_{un}(M) = S - 2^{N_{nonzero}-N_T} + 1.$$

That is, the bigger $N_T$ is, the larger the brute-force space is.

*Imperceptibility and Security* According to the above analysis, the relation between imperceptibility and security is

$$f_{un}(M) = S - 2^{N_{nonzero}-o(I)} + 1.$$

It shows that the higher the imperceptibility I is, the larger the brute-force space is.

## 6 The Perceptual Security

For media encryption, perceptual security [2] is required, which means that the encrypted media data should be intelligible. Generally, the significant part of media data is encrypted in partial encryption. In Kundur's scheme, the DCT coefficients in low frequency band are preferred to be encrypted, and the more the coefficients are encrypted, the higher the perception security is. Fig. 6 shows the relation between the number of encrypted coefficients $N_{en}$ and the media quality (Lena, q=0.5). Thus, in sign encryption, the $N_{nonzero}$ nonzero coefficients are all encrypted. However, in DCT transformation, only sign encryption is not secure enough, for the amplitude is left unchanged. Fig. 7(a) and (b) give the original image, and the encrypted image that is still intelligible. Thus, it is not secure enough for some applications.

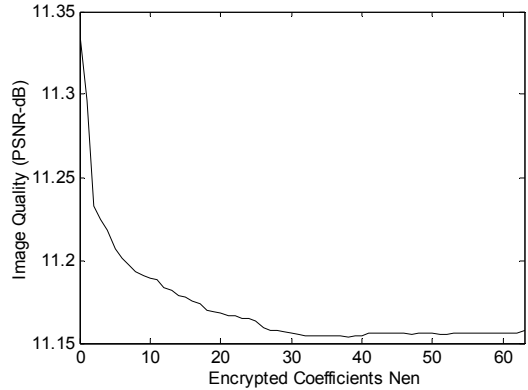

**Fig. 6**. Perceptual Quality and Number of Encrypted Coefficients $N_{en}$

## 7 Means to Improve the Scheme's Performances

According to the above analysis, the security of Kundur's scheme is in close relation with the scheme's robustness and imperceptibility. To keep high robustness, the number of nonzero coefficient $N_{nonzero}$ should be decreased, although it reduces the security in some extent.

To keep good imperceptibility, the threshold coefficient $N_T$ should be increased, although it reduces the number of customers that can be supported. In order to improve the scheme's performances, several means are presented as follows.

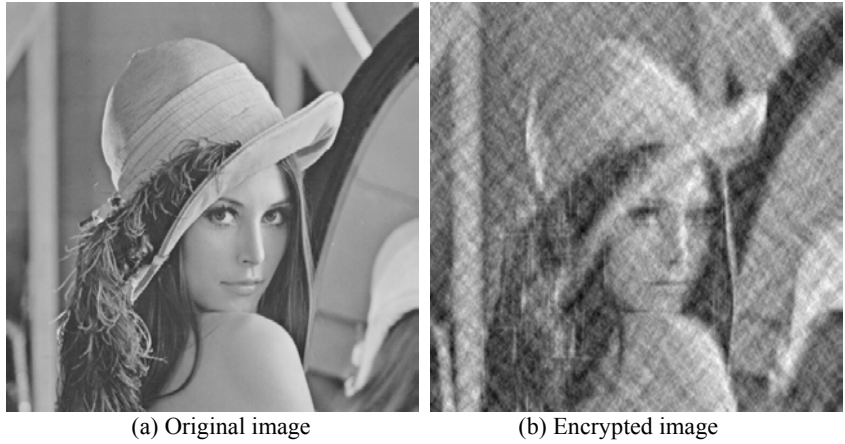

(a) Original image          (b) Encrypted image

**Fig. 7.** Perceptual Security of Sign Encryption

### 7.1 MultiKey Encryption

As is mentioned in Section 3, the brute-force space decreases with rise of the number of the customers that can be supported. To enlarge the brute-force space is a solution to this problem. That is, the reduced brute-force space still satisfies the requirement of practical applications. Thus, multi-key encryption can be used here: media data are partitioned into H parts, and each part is encrypted with a different key. In this case, the brute-force space is

$$\begin{cases} f_{un}(M) = S^H - M + 1 \\ f_{au}(M) = S^H - M \end{cases}.$$

Compared with the previous one, the space is greatly enlarged, and the number of the customer does little effect on the scheme's security.

### 7.2 Selection of $N_T$ and $N_{nonzero}$

The number of $N_{nonzero}$ is in close relation with the security and robustness. Generally, it is computed by setting a suitable quantization factor. According to the required robustness (the supported compression ratio in JPEG/MPEG), the maximal quantization factor can be obtained, which is used to generate $N_{nonzero}$. Generally, under q=0.5, the average value is $N_{nonzero}$=16. Differently, the threshold coefficient $N_T$ determines both the imperceptibility and the number of customers. For $N_T$, a tradeoff should be determined between the imperceptibility and the number of customers. Fig. 8 gives the relation between $N_T$ and the quality of the decrypted image. Generally, the minimal value is $N_T$=8, which keeps the imperceptibility. Thus, the maximal number of customer that the scheme can support is $M=2^8$. Additionally, it is practical to enlarge the number by multi-key method. That is, the image is partitioned into L parts, and each part is decrypted with different mode. Thus, the maximal customer number is changed into $M=2^{8L}$.

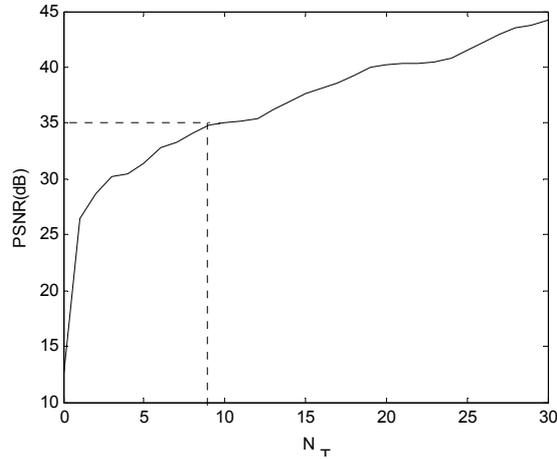

**Fig. 8.** Relation between $N_T$ and the Quality of the Decrypted Image

### 7.3 Improvement of the Perceptual Security

As is mentioned in Section 6, the perceptual security should be improved. Considering that the energy is concentrate on low frequency in DCT transformation, encrypting the coefficients in low frequency causes great blurs to the decoded media data. Here, DC encryption is inserted into Kundur's scheme, which encrypts DC coefficients besides their signs. The encryption operation should be applied following quantization process, in order to make the media data decrypted correctly. Fig. 9 shows the results of image encryption. As can be seen, the encrypted image (b) is unintelligible, which shows that the improved scheme is more secure in perception.

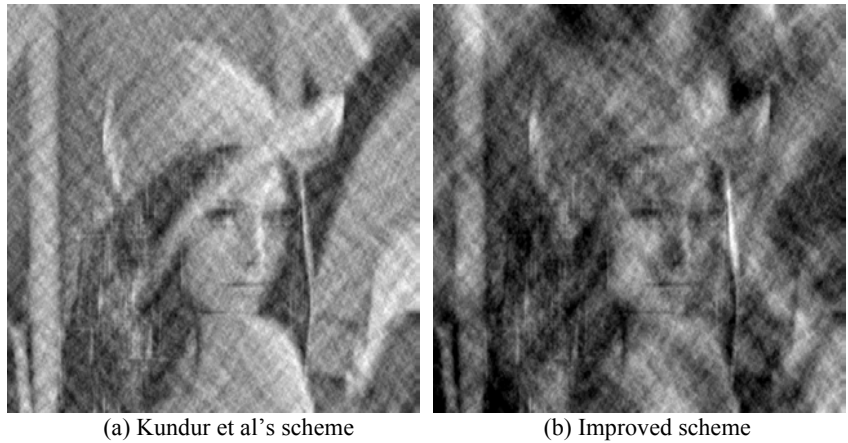

(a) Kundur et al's scheme          (b) Improved scheme

**Fig. 9.** Perceptual Security of Image Encryption

# 8 Conclusions and Future Work

In this paper, the performances of Kundur et al's JFD scheme are analyzed. It is pointed out that the security often contradicts with the robustness, the imperceptibility often contradicts with the maximal number of the supported customers, and the security often decreases with the rise of the maximal number of the supported customers. In order to obtain good tradeoff between security, robustness and imperceptibility, some means are proposed, including multi-key encryption, multi-key decryption and DC encryption. It is expected to provide valuable information for designing a JFD scheme. The analysis method and the proposed means may also be used to evaluate other JFD schemes, which is our future work.